\documentclass[aps,prb,citeautoscript,twocolumn,superscriptaddress,bibnotes]{revtex4-1}
\usepackage{float}
\usepackage{physics}
\usepackage{graphicx}
\usepackage{xcolor}
\usepackage{amsmath}
\usepackage{textcomp}

\renewcommand{\figurename}{FIG.}
\makeatletter
\renewcommand*{\fnum@figure}{{\normalfont\bfseries \figurename~\thefigure}}
\renewcommand*{\@caption@fignum@sep}{ $~$}

\renewcommand{\tablename}{Table}
\makeatletter
\renewcommand*{\fnum@table}{{\normalfont\bfseries \tablename~\thetable}}

\usepackage[pdftex,bookmarks=true,bookmarksopen,bookmarksnumbered,
                colorlinks,
                linkcolor=blue,
                citecolor=blue]{hyperref}

\usepackage{float}
\usepackage{newfloat}
\usepackage{standalone}
\usepackage{booktabs}
\usepackage[geometry]{ifsym}

\graphicspath{{paper_figures/}}

\newcommand{\ktao}{\texorpdfstring{KTaO\textsubscript{3}~}{KTaO3}}

\newcommand{\nv}{\texorpdfstring{NV\textsuperscript{-}~}{NV-~}}

\newcommand{\Bzero}{\textit{B}\textsubscript{0}~}
\newcommand{\Bone}{\texorpdfstring{\textit{B}\textsubscript{1}~}{B1~}}

\usepackage{lineno}
\begin{document}

\title{Fast coherent control of an \nv spin ensemble using a \ktao dielectric resonator at cryogenic temperatures}

\author{Hyma H. Vallabhapurapu}
\email[]{h.vallabhapurapu@student.unsw.edu.au}
\affiliation{
 Centre for Quantum Computation and Communication Technology,
 School of Electrical Engineering and Telecommunications,
 The University of New South Wales, Sydney, NSW 2052, Australia
}

\author{James P. Slack-Smith}
\affiliation{
 School of Electrical Engineering and Telecommunications,
 The University of New South Wales, Sydney, NSW 2052, Australia
}

\author{Vikas K. Sewani}
\affiliation{
 Centre for Quantum Computation and Communication Technology,
 School of Electrical Engineering and Telecommunications,
 The University of New South Wales, Sydney, NSW 2052, Australia
}

\author{Chris Adambukulam}
\affiliation{
 School of Electrical Engineering and Telecommunications,
 The University of New South Wales, Sydney, NSW 2052, Australia
}

\author{Andrea Morello}
\affiliation{
 Centre for Quantum Computation and Communication Technology,
 School of Electrical Engineering and Telecommunications,
 The University of New South Wales, Sydney, NSW 2052, Australia
}

\author{Jarryd J. Pla}
\affiliation{
 School of Electrical Engineering and Telecommunications,
 The University of New South Wales, Sydney, NSW 2052, Australia
}

\author{Arne Laucht}
\email[]{a.laucht@unsw.edu.au}
\affiliation{
 Centre for Quantum Computation and Communication Technology,
 School of Electrical Engineering and Telecommunications,
 The University of New South Wales, Sydney, NSW 2052, Australia
}

\begin{abstract}
Microwave delivery to samples in a cryogenic environment can pose experimental challenges such as restricting optical access, space constraints and heat generation. Moreover, existing solutions that overcome various experimental restrictions do not necessarily provide a large, homogeneous oscillating magnetic field over macroscopic lengthscales, which is required for control of spin ensembles or fast gate operations in scaled-up quantum computing implementations. Here we show fast and coherent control of a negatively charged nitrogen vacancy spin ensemble by taking advantage of the high permittivity of a \ktao dielectric resonator at cryogenic temperatures. We achieve Rabi frequencies of up to 48~MHz, with a total field-to-power conversion factor $C_{\rm P}= ~$9.66~mT/$\sqrt{\rm W}$ ($\approx191$~MHz\textsubscript{Rabi}/$\sqrt{\rm W}$). We use the nitrogen vacancy center spin ensemble to probe the quality factor, the coherent enhancement, and the spatial distribution of the magnetic field inside the diamond sample. The key advantages of the dielectric resonator utilised in this work are: ease of assembly, in-situ tuneability, a high magnetic field conversion efficiency, a low volume footprint, and optical transparency. This makes \ktao dielectric resonators a promising platform for the delivery of microwave fields for the control of spins in various materials at cryogenic temperatures.
\end{abstract}

\maketitle

\section{\label{sec:introduction} Introduction}

\begin{figure*}[t]
\centering
\includegraphics{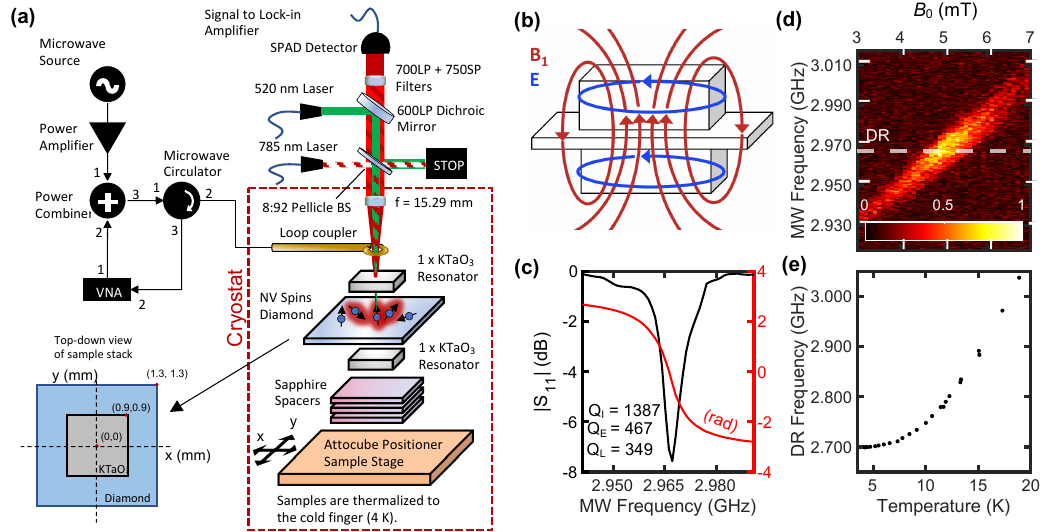}
\caption{Setup and ODMR response.
\textbf{(a)} Overview of the measurement setup showing the schematic of the microwave circuit, exploded view of the optical setup and sample mounting. The samples are mounted on an Attocube piezoelectric stage, allowing the samples to be re-positioned inside the cryostat.
\textbf{(b)} Sketch of the magnetic (red) and electric (blue) field lines corresponding to the $\rm{TE}_{11\delta{}}$ excitation mode of the dielectric resonator.
\textbf{(c)} Reflection magnitude and phase signals of the overcoupled \ktao dielectric resonator, when tuned through a fixed microwave reference frequency at $2.967$~GHz.
\textbf{(d)} ODMR signal as a function of microwave frequency and magnetic field $B_0$. The white dashed line indicates the dielectric resonator frequency. 
\textbf{(e)} Measured temperature dependence of the dielectric resonator (DR) frequency in the cryostat.} 
\label{fig:sample_setup_diagram}
\end{figure*}

\begin{figure*}[ht]
 \centering
\includegraphics[keepaspectratio]{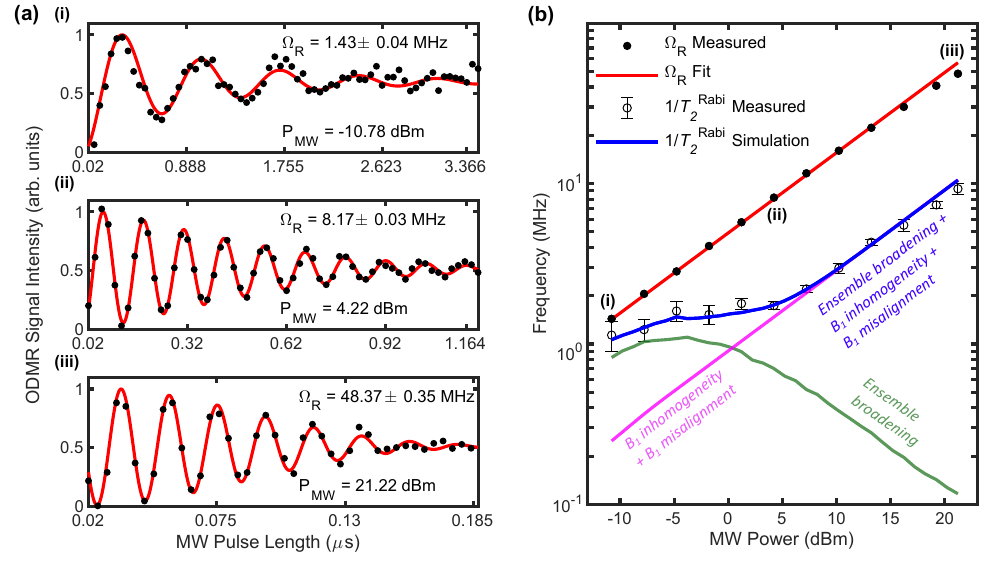}
\caption{\Bone conversion factor and Rabi decay rate.
\textbf{(a)} Selected Rabi oscillation measurements at three microwave powers, specified at the loop coupler. The red lines are fits to the measured data with a decaying sinusoidal function including an exponent $n$.
\textbf{(b)} The measured Rabi frequency is shown to double for +6 dB increments of microwave power, as expected (the red fit excludes the last three data points due to amplifier saturation). The trend of the 1/${T_{2}}$\textsuperscript{Rabi} decay rate (blue line) can be explained by power broadening. This leads, in combination with the spectral distribution of the ensemble (green line) and the \Bone inhomogeneity within the detection volume (magenta line), to the observed S-bend. }
 \label{fig:rabi_power_decay_simulation}
\end{figure*}

Fast spin control -- much faster than the coherence time -- is essential for achieving truthful and high-fidelity qubit gate operations \cite{DiVincenzo2000, Nielsen2010}. While dynamical decoupling and error-correction can be used to somewhat mitigate decoherence effects, they only work once a certain fidelity threshold has already been met \cite{yang2011preserving,lidar2013quantum}. In the context of spin qubit control, microwave (MW) resonators can allow for an efficient interaction with the spin qubits of interest \cite{LeFloch2014,Vahapoglueabg9158,Ebel2020,eisenach2021cavity}. In particular, the enhancement of the oscillating magnetic field ($B_1$) offered by such a resonator can result in high-frequency Rabi oscillations of the spins, and thus fast gate operations, while keeping the input power at a moderate level. A prominent figure of merit is the \Bone field-to-power conversion factor $C_{\rm P}$, which can be accurately estimated by measuring the Rabi frequency $\Omega_{\rm R}$ of the spins through the relationship $\Omega_{\rm R} = \tfrac{1}{\sqrt{2}}\gamma_{\rm e}B_1 = \tfrac{1}{\sqrt{2}}\gamma_{\rm e}C_{\rm P}\sqrt{P_{\rm MW}}$ (for $m_{s} = 1$), where $C_{\rm P}=B_1/\sqrt{P_{\rm MW}}$ and $\gamma_{e}$ is the electron gyromagnetic ratio.

For cryogenic environments, a suitable MW field delivery method designed for optically-active spin defects is yet to be realised, especially one that is homogeneous enough for ensemble measurements. Some favourable attributes of an ideal MW delivery apparatus for such an application include optical transparency, high \Bone conversion factor at cryogenic temperatures, in-situ resonance frequency tuneability, and simple design and implementation. Existing solutions include conventional antenna and resonator designs \cite{Chipaux2015,Bayat2014, Sasaki2016, Herrmann2016, Mrozek2015, Eisenach2018, Mett2019, Yaroshenko2020, wang2020} and spin-wave mediated methods \cite{Andrich2017}, but lack in one or more of the aforementioned characteristics. 

Critically, high speed quantum gate operations require fast Rabi oscillations of the spin qubit, ultimately benefiting from a high $C_{\rm P}$. Typically, such high $C_{\rm P}$ can be achieved using nanoscale, lithographically defined, on-chip antennas \cite{Dehollain2013} and superconducting resonators \cite{Bienfait2016a}. However, these structures require complex on-chip fabrication, and their use is often restricted to spins located in close proximity to the antenna structure. Despite benefiting from a high $C_{\rm P}$, practically achievable Rabi frequencies range from 100 kHz to 3 MHz due to various damage thresholds and superconductivity limitations including critical current limits. However, Rabi oscillations as fast as 440 MHz have also been achieved outside a cryostat \cite{Fuchs2009} using similar structures. The highest reported $C_{\rm P}$ for macroscopic MW resonators in literature are 1.19 mT/$\sqrt{\rm W}$ (23.60 MHz\textsubscript{Rabi}/$\sqrt{\rm W}$) \cite{Bayat2014, Yaroshenko2020} for planar split-ring resonators, 1.17 mT/$\sqrt{\rm W}$ (23.16 MHz\textsubscript{Rabi}/$\sqrt{\rm W}$) for a rutile dielectric loop-gap resonator~\cite{Mett2019} and 3.2 mT/$\sqrt{\rm W}$ (89.60 MHz\textsubscript{Rabi}/$\sqrt{\rm W}$) for a microhelix resonator \cite{Sidabraseaay1394}.

Dielectric resonators in particular have played a prominent role in spin resonance experiments \cite{Annino1999} with regards to enhancement of the MW excitation field \cite{Walsh1986,Geifman2006} and demonstration of the spin-cavity coupling dynamics \cite{Abe2011,LeFloch2016,Breeze2018,Ebel2020,eisenach2021cavity}. \ktao is a quantum paraelectric material possessing low MW losses (tan $\delta\sim 10^{-4}-10^{-5}$) and high permittivity ($\epsilon_{\rm r}\approx 4300$ for $T<10$~K), making it an encouraging material for delivering high MW fields at cryogenic temperatures\cite{Geyer2005}, in contrast to commonly used dielectrics such as rutile and sapphire \cite{Sabisky1962,Krupka1999} that possess $\epsilon_{\rm r}$ up to two orders of magnitude lower. For a given frequency, the required dimensions of the dielectric resonator scales as a function of $\sqrt{\epsilon_{\rm r}}$ \cite{Walsh1986}. A significant reduction in size is expected of \ktao resonators compared to sapphire or rutile resonators at cryogenic temperatures. For operation in the gigahertz range this results in convenient millimetre dimensions for \ktao resonators. Further, due to its good optical properties and high refractive index ($n=2.3$ for $\lambda=500$~nm at 300 K), \ktao has been used as a solid immersion lens \cite{Shinoda2006,KazuoFujiura2007}, showing great potential for application as a dual-purpose MW resonator that can be used for boosting photon collection efficiency and MW field strength simultaneously for optically-detected magnetic resonance of spin defects.

In the past, \ktao dielectric resonators have been used to demonstrate enhanced electron paramagnetic resonance (EPR) sensitivity \cite{Geifman2000,Blank2003,Geifman2006} (measured $C_{\rm P} = 1.3~\rm{mT}/\sqrt{\rm W}$ at 77 K), and for incoherent spin manipulation of a quantum dot spin qubit in a SiMOS device \cite{Vahapoglueabg9158}. However, little work has been done demonstrating coherent control of a solid-state spin -- single or ensemble -- at cryogenic temperatures.  

Here we demonstrate the use of a \ktao dielectric resonator with internal quality factor $Q_{\rm I}\approx1387$ (limited by dissipation in the diamond sample) and $>100$~MHz in-situ tuneability to deliver high-amplitude oscillating \Bone fields to a NV\textsuperscript{-} spin ensemble in diamond. This allows us to drive fast and coherent Rabi oscillations up to $\Omega_{\rm R}=48$~MHz, and conversely use the \nv spin ensemble to probe the resonator characteristics. We measure a conversion factor $C_{\rm P}$ of 9.66~mT/$\sqrt{\rm W}$ ($\approx191$~MHz\textsubscript{Rabi}/$\sqrt{\rm W}$) and a $\sim 7$\% spatial \Bone inhomogeneity over the $0.4$~mm central region. By performing $T_2^{\rm Rabi}$ and $T_2^{\rm Hahn}$ echo measurements, we identify the dominant limitation to the control fidelity to be the $B_1$ inhomogeneity within the signal collection volume and a slight misalignment of the $B_1$ field with respect to the crystallographic axis of the diamond sample.

We choose an ensemble of negatively charged nitrogen vacancy (NV\textsuperscript{-}) centres in diamond \cite{DOHERTY20131} to characterize the dielectric resonator due to the high spin-dependent photoluminescence contrast down to cryogenic temperatures. \nv centres in diamond have a wide range of applications in the field of magnetometry \cite{Rondin_2014}. The spin ensemble conveniently allows us to spatially map the \Bone field of the dielectric resonator within the diamond; similar mapping of the $B_1$ field of other resonators has been demonstrated in literature\cite{Appel2015,horsley,Yang2019, Sasaki2016,Yaroshenko2020}. Finally, colour centres in diamond provide a promising qubit platform for practical implementations of future quantum networks and computers \cite{Nemoto2016,Bhaskar2020,Sukachev2017,Wrachtrup2006}-- thus it is useful to develop such methods to demonstrate compatibility with diamond-based spins.

\section{Results}
\subsection{Experimental setup and methods}
 \textbf{Setup and samples.} Our measurement setup is illustrated in Fig.~\ref{fig:sample_setup_diagram}(a). It consists of MW circuitry to measure the input port reflection parameter ($S_{11}$) of the coaxial loop coupler using a vector network analyser (VNA). This circuitry allows the resonator to be driven by a MW source while simultaneously monitoring the resonance during measurements. The optical part of the setup is used to optically interact with the NV\textsuperscript{-} spin ensemble. The optical setup delivers a 520 nm green laser for off-resonant excitation of the NV\textsuperscript{-} centers with a focused beam spot diameter of $\sim$ 5 $\mu$m (see Table~\ref{tab:laser_params}). The resulting emission from the NV\textsuperscript{-} is then collected by the same optical setup via a different optical path using an appropriate dichroic mirror. The optical setup also delivers an $785$~nm infrared laser for in-situ, thermal tuning and stabilization of the dielectric resonator (see Fig.~\ref{fig:infrared_control}). A lens with a long focal length of 15.29~mm is used to focus all the lasers into the diamond sample, while also collecting the NV\textsuperscript{-} photoluminescence.

 A $2.6\times2.6\times0.25$~mm$^3$ diamond sample containing an \nv spin density of $\sim 6\times10^{11}$ spins/$\rm mm^{3}$ (grown by chemical vapour deposition with $\langle100\rangle$ orientation) is sandwiched between two \ktao prisms ($1.8\times1.8\times0.5$~mm$^3$) that form the dielectric resonator (see Fig.~\ref{fig:sample_stack}(a) in Appendix). The samples are placed on a piezoelectric sample stage (Attocube), which allows re-positioning of the samples inside the cryostat. Finally, a loop coupler positioned above the samples is used to couple to the dielectric resonator inside the cryostat such that the resonator is operating in the overcoupled regime (see Figs.~\ref{fig:sample_stack}(b) and \ref{fig:coupling_regimes}). The coupling of the loop coupler to the dielectric resonator is shown in Fig.~\ref{fig:sample_setup_diagram}(c), with loaded and internal quality factor fits of $Q_{L} = 349\pm5$ and $Q_{I} = 1388\pm6$, respectively. The fundamental $\rm{TE}_{11\delta{}}$-like mode of the \ktao dielectric resonator is used for all measurements, illustrated in  Fig. \ref{fig:sample_setup_diagram}(b). The single spin to cavity coupling strength is $g_{\rm s}/2\pi \approx 1$ Hz, approximately given by $\gamma_{e}\sqrt{\mu_{0}\hbar{}w_{0}/(2V_{\rm m})}$ \cite{Breeze2018}  for a simulated mode volume $V_{\rm m} = 0.98~{\rm mm}^3$ (see Appendix Sec. 2). The parameters $\mu_{0}$, $\hbar{}$ and $w_{0}$ being the permeability of free space, reduced Plank's constant and resonator frequency in units of rad/s respectively.

  \textbf{ODMR measurements.} All spin experiments are performed using pulsed optically detected magnetic resonance (ODMR) to observe the absolute change in photoluminescence as a result of transferring the spin population from the $\ket{0}$ initial state to the $\ket{+1}$ state \cite{Zhang2018,Sewani2020}, using a lock-in amplifier. A superconducting magnet is used to sweep the static \Bzero field; it is set to persistence mode when the $B_0$ field is required to be fixed in order to ensure $B_0$ stability. All error bars for measurement fits are given as 95\% confidence intervals.
  
  \textbf{Resonator tuning.} The resonator is continuously tuned to 2.967 GHz or another predetermined frequency using the 785 nm laser (see Fig.~\ref{fig:infrared_control}), corresponding to a local resonator temperature of 17~K, estimated using the measured frequency versus cryostat temperature curve shown in Fig.~\ref{fig:sample_setup_diagram}(e). The temporal stability of the resonance frequency is recorded throughout the measurements and is shown in Fig.~\ref{fig:histograms}. Apart from the thermal tuning technique, the dielectric resonator is further stabilized by employing pulse compensation wherein the total duty cycle of the laser and MW pulses are kept constant in the ODMR pulse sequences (see Fig.~\ref{fig:mwpulse_control}). 

 Additional details of the experimental setup can be found in Sec.~\ref{Supp_Sample} of the Appendix.

\subsection{ODMR signal enhancement}
\label{odmr_enhancement}

We start by performing incoherent ODMR measurements using the measurement setup shown in Fig.~\ref{fig:sample_setup_diagram}(a) to observe the ODMR signal enhancement. In Fig.~\ref{fig:sample_setup_diagram}(d), we sweep \Bzero to tune the \nv spin transitions through the MW dielectric resonance. The measurement is performed using low MW power ($-11$~dBm) at the loop coupler and long MW pulses (40~$\mu$s) to reduce power broadening and to ensure that the spins are incoherently driven. The result clearly shows the increase in the spin-dependent fluorescence signal when the \nv frequency is in the vicinity of the cavity resonance. Integration of the ODMR signal for the on-resonance and off-resonance cases presents an ODMR signal enhancement factor of 1.7.

\subsection{\Bone field conversion}

Next, we extract the power-to-field conversion factor $C_{\rm P}$ by measuring the Rabi frequency $\Omega_{\rm R}$ of the spin ensemble as a function of applied MW power $P_{\rm MW}$ over three orders of magnitude. The measurements are performed with the MW source, \nv and resonator frequencies in resonance at 2.967~GHz, as shown in Fig.~\ref{fig:sample_setup_diagram}(c), corresponding to $\Bzero = 5$~mT. We plot the Rabi oscillations for selected MW powers in Fig.~\ref{fig:rabi_power_decay_simulation}(a). We fit the oscillations to decaying sinusoids (red lines) and plot the extracted Rabi frequencies ($\Omega_{\rm R}$) and Rabi decay rates ($1/{T_{2}}$\textsuperscript{\rm Rabi}) in Fig.~\ref{fig:rabi_power_decay_simulation}(b). 
The $\Omega_{\rm R}$ data (filled circles) in Fig.~\ref{fig:rabi_power_decay_simulation}(a) is then fit to a linear function of $\sqrt{\rm W}$ with a gradient of $C_{\rm P}=156\pm2$~MHz\textsubscript{Rabi}/${\sqrt{\rm W}}$, representing the measured Rabi frequency conversion factor. After compensating for the oscillating field within the rotating wave approximation and the misalignment between the \Bone direction and the \nv quantization axis (see Sec.~\ref{Supp_B1Conv} of the Appendix), we find the total $B_{1}$ conversion factor to be 9.66~mT/$\sqrt{\rm W}$, corresponding to a theoretical Rabi frequency conversion factor of 191 MHz\textsubscript{Rabi}/$\sqrt{\rm W}$. This conversion factor is in good agreement with simulated results (see Appendix Sec.~\ref{cst_sim}), as shown in Fig.~\ref{fig:all_hfield}(c), as well as the theoretically predicted mean value \cite{poole1996electron} 
\begin{equation}
\label{eq:conversion_b1}
    \left\langle C_{\textrm{P}} \right\rangle = \mu_{0}\sqrt{\frac{2Q_{L}}{\mu_{0}V_{m}\omega_0}}\approx \langle7~{\rm mT}/\sqrt{\rm W}\rangle
\end{equation}

The $C_{\rm P}$ measured in this work is at least a factor 7 larger than those reported for dielectric resonators in literature \cite{Blank2003, Mett2019} including previously reported \ktao resonators, and at least a factor of 3 larger than other resonators \cite{Sidabraseaay1394,Bayat2014,Yaroshenko2020}. This is a considerable improvement, as to achieve the same \Bone fields shown in this work using other resonators would typically require at least a +9 dB addition of MW power, which can be difficult to achieve in millikelvin cryostats.

\Bone conversion factors up to an order of magnitude higher should be achievable with higher quality factors of the dielectric resonator as can be predicted from Eq.~\ref{eq:conversion_b1}, for example by using an electronic grade diamond sample that exhibits lower loss (see Table~\ref{tab:diamond_comparison} in Appendix). High quality factors are expected to proportionally increase the ringdown time of the resonator. It is trivial to see that the final spin state is directly dependent on the integral of the applied MW pulse signal. Therefore, a short square shaped MW pulse is approximately equivalent to a resulting bandwidth limited MW pulse, given that the spins are not reinitialised before the ringdown time $\tau = 2\frac{Q_{L}}{\omega_{0}}$. This is appropriately accounted for in the measurements by using a delay time (of order 100 ns) between the microwave pulse and a subsequent laser readout pulse within a relevant pulse sequence. However, this may not be necessary for applications utilizing a `global' MW field \cite{Vahapoglueabg9158}, wherein the spin system can instead be arbitrarily brought into and out of resonance of the MW field for coherent control. Alternatively, shaped MW pulses may be used to achieve ringdown suppression \cite{probst201942}.

The open circles in Fig.~\ref{fig:rabi_power_decay_simulation}(b) depict the corresponding Rabi decay rate $1/{T_{2}}$\textsuperscript{\rm Rabi}. The observed S-bend trend can be explained as follows. At low MW powers, the dominant limitation to $T_{2}^{\rm Rabi}$ is the ensemble broadening given by the hyperfine coupling to the $^{14}$N nuclear spins and the four \nv centre axes that form slightly different angles with respect to $B_0$, leading to the summation of signals from twelve transitions at varying detuning from the excitation frequency. When increasing $P_{\rm MW}$, power broadening improves $T_{2}^{\rm Rabi}$~\cite{Wrachtrup2006} when it becomes greater than the inhomogeneous spin ensemble linewidth, leading to an inflection in the decay rate trend. At even higher $P_{\rm MW}$, we find that $T_{2}^{\rm Rabi}$ is limited by the \Bone inhomogeneity, leading to an increase in $1/{T_{2}}$\textsuperscript{\rm Rabi} that is directly proportional to $\Omega_{\rm R}$. This \Bone inhomogeneity manifests as a combination of the spatial variation of the \Bone field produced by the dielectric resonator within the signal collection volume (limited by the laser focus), and the different effective \Bone strengths corresponding to the four unique \nv orientations within the diamond crystal lattice due to some misalignment with respect to their quantization axes~\cite{Sewani2020}. 

We perform numerical simulations considering the aforementioned effects, with results shown in Fig.~\ref{fig:rabi_power_decay_simulation}(b). The simulated Rabi decay rates (blue line) are in good agreement with the experimental data (open circles). In the same figure we also show the individual contributions of the \Bone inhomogeneity (magenta line) and ensemble broadening (green line) towards the total Rabi decay rate. More information about the simulation model can be found in Sec.~\ref{Supp_Rabi} of the Appendix. A collection of measured and simulated Rabi oscillations can be seen in Fig.~\ref{fig:rabi_collection}.

\subsection{Coherent \Bone enhancement}

\begin{figure}[t]
	\centering
	\includegraphics{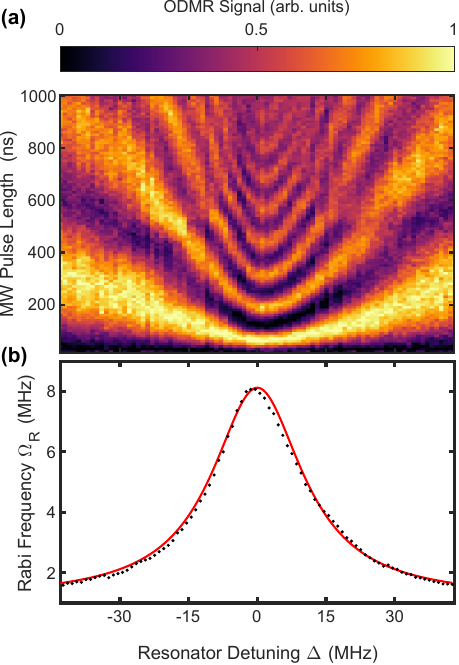}
	\caption{Detuning dependence of Rabi oscillation enhancement.
	\textbf{(a)} Chevron pattern of the Rabi oscillations plotted against a range of \nv -- resonator detunings. The microwave source was kept in resonance with the NV\textsuperscript{-} ODMR frequency throughout this measurement. 
	\textbf{(b)} Corresponding Rabi frequencies fitted to a Lorentzian function. The on-resonance enhancement is calculated as $\approx 7.1 \times$.}
	\label{fig:detuning_chevron}
\end{figure}

In this section we investigate the detuning dependence of the \Bone field enhancement by the dielectric resonator, by measuring the Rabi frequencies for various resonator frequency detunings relative to the \nv and MW source frequency. We therefore fix the MW source to 2.967~GHz and the \Bzero field to $\approx 5$~mT such that the \nv transitions are in resonance with the driving field, and sweep the resonator frequency through various predetermined detuning frequencies using the 785~nm laser. This method ensures that frequency dependent interferences in the cables do not lead to a variation in the MW power arriving at the dielectric resonator. The measurement results in a chevron-like pattern of Rabi oscillations which we plot in Fig.~\ref{fig:detuning_chevron}(a). We extract the resulting Rabi oscillation frequencies $\Omega_{\rm R}$ and plot them as a function of resonator detuning $\Delta$ in Fig.~\ref{fig:detuning_chevron}(b). The red line is a Lorentzian fit to the data.

The linewidth fit of $\Omega^{2}_{\rm R}$ is found to be 15.6 MHz and corresponds to the FWHM linewidth for a 3~dB drop in microwave power. This linewidth is a result of the convolved linewidths of the resonator and that of the ensemble broadened spin resonance. The spin resonance linewidth can be approximated as $7.1\pm1.7$ MHz using the average ODMR linewidth of the detuned ODMR resonance ($B_{0}\approx$~3 mT) from Fig.~\ref{fig:sample_setup_diagram}(d). The bare resonator linewidth is thus simply calculated as the difference $8.5\pm1.7$ MHz, corresponding to a $Q_{L} = 351^{+86}_{-58}$. This is in good agreement with the resonator linewidth and quality factor measured by the VNA, shown in Fig.~\ref{fig:sample_setup_diagram}(c).

On resonance, we observe a coherent enhancement factor of 7.1  for the ODMR signal, higher than the incoherent enhancement shown in Fig.~\ref{fig:sample_setup_diagram}(d). This is due to the fact that $\Omega_{\rm R}$ provides an accurate, direct measurement of $B_1$, while the enhancement of the incoherent ODMR signal is an indirect measurement of a spin ensemble in a mixed state and subject to saturation effects.

In addition to the detuning dependence of the \Bone enhancement, the measurement in Fig.~\ref{fig:detuning_chevron} also demonstrates the resonance frequency tuneability of the dielectric resonator via heating with the 785~nm infrared laser (see also Sec.~\ref{Supp_Sample} of the Appendix for more details on resonator frequency tuning and stabilization). In this measurement, we tune the resonator over a range of $\sim 90$~MHz. Such tuneability allows for some flexibility in the dielectric resonator dimensions and is useful in experiments where the $B_0$ field cannot be adjusted\cite{adambukulam2020ultra}, or where ODMR transitions may be selectively enhanced without needing to adjust the \Bzero field.

\begin{figure}[h!]
	\centering
	\includegraphics{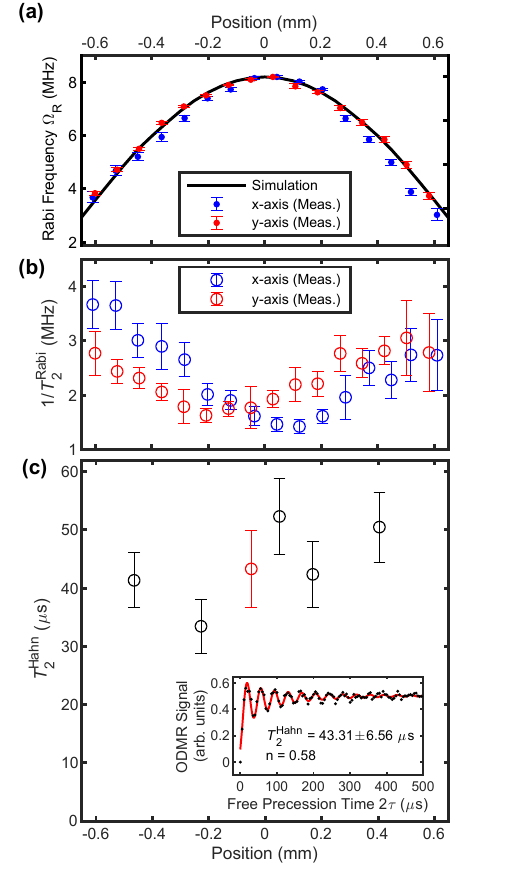}
	\caption{Position dependence and homogeneity of \Bone enhancement.
	\textbf{(a)} Measured and appropriately compensated Rabi frequency $\Omega_{\rm R}$ along the x- and y-axes of the diamond sample (see Eq. \ref{eq:compensation} in Appendix). 
	The black line corresponds to predicted $\Omega_{\rm R}$ values, obtained from modelling the trend of the $B_{1}$ field of the dielectric resonator in CST Microwave Studio. We find good agreement between the experimental data and the CST simulation model.
	\textbf{(b)} 1/${T_{2}}$\textsuperscript{Rabi} decay rates measured and plotted across both axes of the sample. We observe a general increase in  ${T_{2}}$\textsuperscript{Rabi} coherence times towards the centre of the resonator, due to higher \Bone homogeneity. 
	\textbf{(c)} ${T_{2}}$\textsuperscript{Hahn} decay times for positions along the x-axis of the resonator. We do not observe significant variations of the ${T_{2}}$\textsuperscript{Hahn} decay times across the sample. The inset shows a sample Hahn echo measurement for a position close to the centre of the resonator, corresponding to the data point highlighted in red in the main panel. Error bars represent 95\% confidence intervals for fits.}
	\label{fig:position_coherence}
\end{figure}

\subsection{Spatial \Bone field distribution and coherence}

We map the \Bone distribution along the x- and y-axes by measuring the corresponding Rabi oscillation frequencies $\Omega_{\rm R}$ and their decay times $T_2^{\rm Rabi}$ at various positions across the sample for a fixed MW excitation power. Since the sample is re-positioned (using the Attocube piezoelectric positioner) to obtain this measurement, the coupling between the loop coupler and the dielectric resonator is expected to change. We numerically compensate the measured Rabi frequencies by accounting for the reflection magnitude ($S_{11}$) as measured by the VNA (more details in Sec.~\ref{Supp_Rabi} of the Appendix). We plot the compensated $\Omega_{\rm R}$ and Rabi decay rates $1/{T_2^{\rm Rabi}}$ along the x- and y-axes in Figs.~\ref{fig:position_coherence}(a) and (b), respectively. $\Omega_{\rm R}$ varies between $8$~MHz at the centre and about $5$~MHz when the sample is moved $0.5$~mm in either direction. Quite remarkably, over the $0.4$~mm centre region (i.e., $-0.2$~mm~$<$~x,y~$<+0.2$~mm), $\Omega_{\rm R}$ only decreases by $\sim 7$\%. 
The $1/{T_2^{\rm Rabi}}$ decay rates are found to be generally lower towards the centre of the resonator, due to the better \Bone homogeneity in the x-y plane.

To obtain further proof that ${T_2^{\rm Rabi}}$ is limited by the increase in \Bone inhomogeneity towards the edges of the sample, we perform Hahn echo measurements at different positions along the x-axis (see Fig.~\ref{fig:hahn_analysis_plot_all}). In a Hahn echo sequence, constant $B_0$ inhomogeneities are refocused and only noise with frequencies at the measurement time scale matters. Furthermore, variations in $B_1$ amplitude only affect the measurement contrast and not the $T_2^{\rm Hahn}$ decay time. The data is fit to the appropriate exponential decay function\cite{Childress2006} including the exponent $n$ as an additional fit parameter. The extracted $T_2^{\rm Hahn}$ decay times are all around $35-55$~$\mu$s (see Fig.~\ref{fig:position_coherence}(c)), in accordance with literature values for our sample type, temperature and $B_0$ misalignment, thus likely limited by noise processes within the diamond sample\cite{Stanwix2010,Takahashi2008,bauch}. Overall, the results show no significant variation across the sample, suggesting little variation of dynamic noise across the dielectric resonator and confirming our assumption that ${T_2^{\rm Rabi}}$ is limited by the inhomogeneity in $B_1$.

\section{Conclusion}

We demonstrate fast coherent control and ODMR measurements of an \nv centre ensemble, utilizing the appreciable conversion factor $C_{\rm P}$ provided by the \ktao dielectric resonator. The simple assembly of the dielectric resonator with the diamond sample, its in-situ tuneability, low volume footprint, and optical transparency have been presented as key advantages in this work. 

We measure Rabi frequencies of up to $48$~MHz and a total \Bone field conversion factor of 9.66~mT/$\sqrt{\rm W}$ ($\approx191$~MHz\textsubscript{Rabi}/$\sqrt{\rm W}$), which shows a  at least a factor of 7 improvement in $C_{p}$ over similar dielectric resonators in literature. An even higher \Bone conversion factor should be possible when using a diamond sample with less dissipation, as a consequence of the higher resulting resonator quality factor. We find that the dominant limiting factor for the Rabi coherence time is the inhomogeneity of the \Bone field produced by the resonator within the signal collection volume, and the effective \Bone inhomogeneity due the \Bone field misalignment with the four possible NV\textsuperscript{-} axes in the diamond. Both these effects are completely inconsequential for measurements on single spins.

The large, homogeneous and long range distribution of the $B_{1}$ fields of the dielectric resonator demonstrated in this work may find useful applications in magnetometry with ensembles of \nv defects\cite{Rondin_2014}, the `global control' scheme of spin qubits \cite{Vahapoglu2020}, and would also benefit `indirect control' nuclear spin techniques \cite{Khaneja2007,swathi} in suitable hyperfine coupled spin systems wherein the electron Rabi frequencies are generally required to be higher than the hyperfine strength (in the range of MHz).

We therefore find \ktao dielectric resonators suitable for MW-enhanced coherent control of optically active spins at cryogenic temperatures, with the exciting prospect of simultaneous application as a solid immersion lens. 

\section{Acknowledgements}
We would like to thank Mark Johnson and Irene Fernandez de Fuentes for assistance with the cryogenic measurements in liquid helium. We acknowledge funds from the Australian Research Council via CE170100012, and J.J.P. via DE190101397. H.V. and J.P.S.-S. acknowledge support from the Sydney Quantum Academy. C.A. and A.L. acknowledge support from the University of New South Wales Scientia program.

\appendix

\renewcommand{\figurename}{FIG.}

\renewcommand{\tablename}{Table}

\section*{Appendix}

\subsection{Samples, apparatus and methods}\label{Supp_Sample}

\begin{figure*}[htb]
	\centering
	\includegraphics{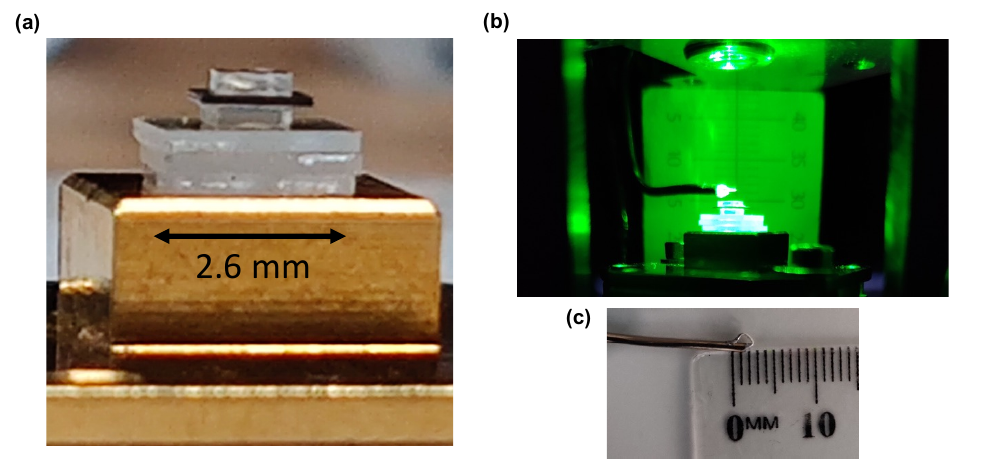}
	\caption{Setup photographs.
	\textbf{(a)} Sample stack with the copper stage and the three sapphire spacers below the \ktao dielectric resonator -- \nv diamond sample sandwich. The sample stack is held together with minimal vacuum grease.
	\textbf{(b)} Sample stack mounted on the dilution refrigerator cold finger atop an Attocube positioner. The coaxial loop coupler is positioned roughly 1.5 mm above the topmost \ktao dielectric resonator. 
	\textbf{(c)} Photo of the coaxial loop coupler showing the size of the looped inner conductor.}
	\label{fig:sample_stack}
\end{figure*}

A $\langle100\rangle$-oriented, $2.6\times2.6\times0.25$~mm$^3$ diamond sample, grown by chemical vapour deposition \cite{E6}, and containing a dilute \nv ensemble, is used to probe the MW field strength and distribution of the resonator. The \nv density was estimated based on comparing the photoluminescence intensity under 520~nm excitation against a sample with a known \nv density. The resonator structure is arranged out of two $1.8\times1.8\times0.5$~mm$^3$ \ktao prisms (manufactured by AWI Industries), with the diamond sample sandwiched in between, and is designed to resonate at approximately 2.7~GHz at 4~K. This sandwich arrangement is similar to a split cylindrical dielectric resonator \cite{Janezic1999}, and helps to create a more homogeneous field distribution along the z-axis as confirmed by simulations [see Figs.~\ref{fig:all_hfield} and \ref{fig:laser_bfield}(b)]. Three sapphire spacers of 0.5~mm thickness are used to distance the resonators from the conductive surface of the sample stage. The sample stack is held together with a minimal amount of vacuum grease on the sides and is mounted on top of an Attocube sample stage which allows positioning of the sample stack along the x-, y- and z-axis. A photo of the sample stack is shown in Fig.~\ref{fig:sample_stack}(a). The sapphire spacers reduce MW losses in the dielectric resonator by physically separating the resonator from the conductive surface of the sample stage. It may be noted that the presence of the sapphire spacers may negatively impact the thermal conduction of heat away from the sample stack diamond and dielectric resonator samples. The optimum thickness of the spacers can be found from a suitable simulation model.

\begin{figure*}[htb]
	\centering
	\includegraphics{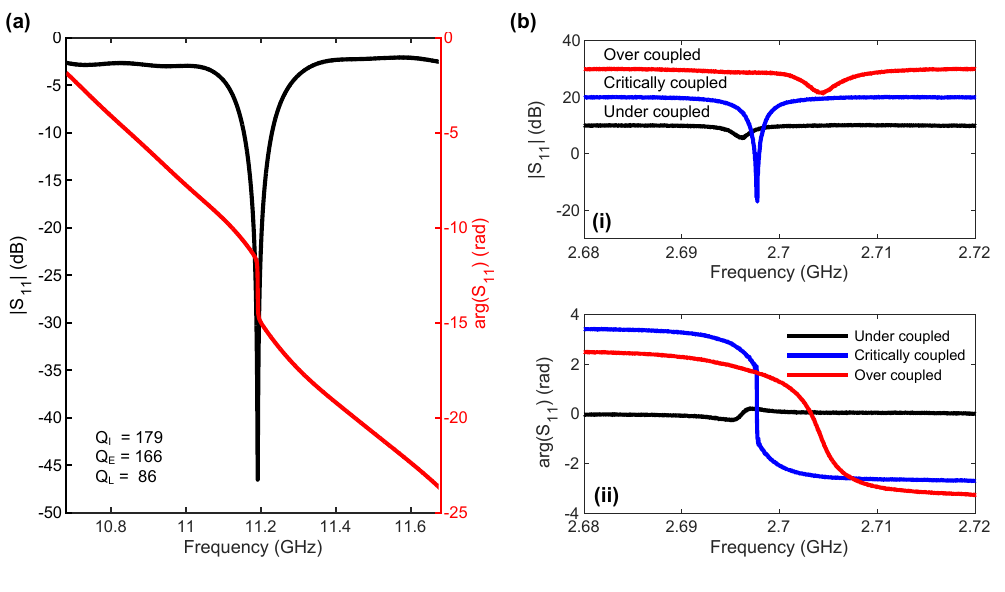}
	\caption{Loop coupler -- resonator coupling.
	\textbf{(a)} VNA measurement of the dielectric resonator stack  at room temperature showing critical coupling. 
	\textbf{(b)} VNA measurements at 4 K showing the S11 magnitude (top panel) and phase (bottom panel) for various coupling regimes of the dielectric resonator to the coaxial loop antenna. The coupling strength can be modified by adjusting the sample z-position. At critical coupling, $Q_{\rm I} = 1275\pm12$, $Q_{\rm E} = 1328\pm5$, $Q_{\rm L} = 650\pm4$. For the over coupled case the values are $Q_{\rm I} = 1414\pm5$, $Q_{\rm E} = 591\pm8$, $Q_{\rm L} = 416\pm7$. The quality factors are extracted using a resonance fitting script \cite{Probst2015}. We attribute the low internal quality factor to the lossy \nv CVD diamond sample (see also Table~\ref{tab:diamond_comparison}).}
	\label{fig:coupling_regimes}
\end{figure*}

The resonator is excited using a coaxial loop coupler that is positioned $\sim 1.5$~mm above the resonator stack [see Fig.~\ref{fig:sample_stack}(b)]. The coupling strength to the coaxial loop coupler can be modified by adjusting the sample z-position with the Attocube stage.
Appropriate coupling to the resonator is ensured by measuring the input port reflection parameter ($S_{11}$) of the loop coupler using a vector network analyser (VNA). At room temperature, we measure the expected reflection dip at $11.2$~GHz due to the much lower permittivity $\epsilon_{\rm r}$ at this temperature. Fig.~\ref{fig:coupling_regimes}(a) shows a measurement at critical coupling. At 4~K, the resonance shifts to $\sim2.7$~GHz, and we demonstrate the various coupling regimes measured with the VNA in Fig.~\ref{fig:coupling_regimes}(b).

The MW circuit consists of a MW source, an amplifier, a VNA, a power combiner and a circulator. The circulator is not strictly necessary for our experiments. However, it is used to prevent damage to our MW equipment due to any high power reflections returning from the cryostat. This is especially important when using a high gain and high power MW amplifier to deliver MW power to a shorted loop coupler. In this way, the circulator directs any reflected MW power through attenuators into port 2 of the VNA. We operate the VNA in $S_{21}$ mode accordingly (this still constitutes a $S_{11}$ reflection measurement on the loop coupler), with a probing power of $\sim$ -40 dBm calculated at the loop coupler. The MW source output is pulse modulated to allow for pulsed spin control. The arrangement of the MW equipment using a power combiner allows for the $S_{11}$ to be measured while performing the coherent spin control measurements. This allows the resonator to be dynamically tuned during the spin measurements using the 785~nm laser for heating and the VNA measurement as a feedback input. 

\begin{figure*}[htb]
	\centering
	\includegraphics{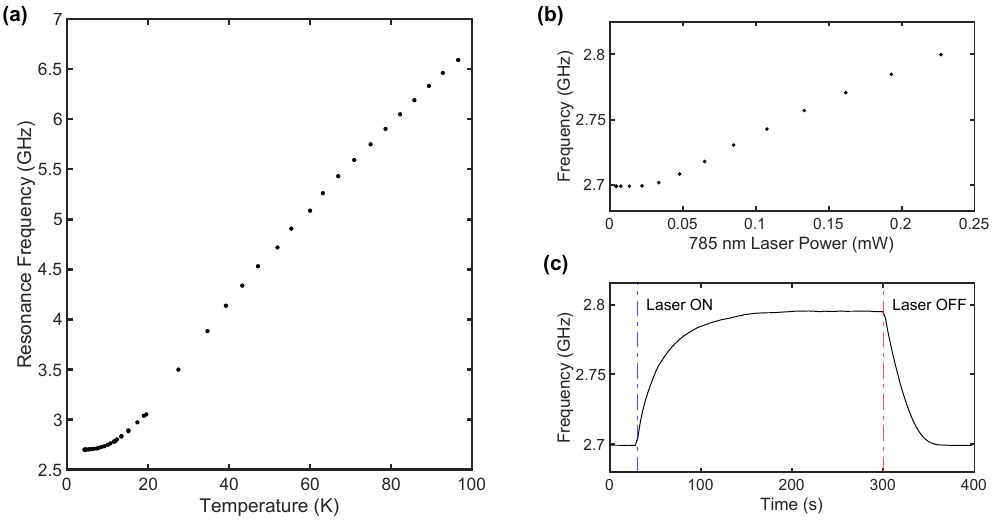}
	\caption{Dielectric resonator frequency tuning.
	\textbf{(a)} Dielectric resonator frequency measured as a function of cryostat temperature. 
	\textbf{(b)} Dielectric resonator frequency as a function of infrared laser power as measured at the optical window into the cryostat. 
	\textbf{(c)} Temporal step response of dielectric resonator frequency after infrared laser modulation.}
	\label{fig:infrared_control}
\end{figure*}

Although the \ktao resonance frequency has been shown to be tuneable via an application of a suitable electrical field\cite{ang}, such a tuning mechanism would require conductor traces to be patterned on the resonator structure or the diamond sample. For the sake of simplicity the dielectric resonator here is tuned thermally by taking advantage of the temperature sensitive permittivity of \ktao (see Fig.~\ref{fig:infrared_control}). The resonator is tuned to the desired frequency via a simple software implemented control algorithm with suitably selected proportional and derivative terms through trial and error. This allows the dielectric resonator temperature to be controlled by heating the sample stack with a low power 785~nm laser. Inspection of Figs. \ref{fig:coupling_regimes}(b)(i) and \ref{fig:sample_setup_diagram}(c) shows that the loaded quality factor of the dielectric resonator does not significantly change when the frequency of the dielectric resonator is shifted from 2.70 GHz ($Q_{\rm L}=416$ corresponding to 4 K) to 2.96 GHz ($Q_{\rm L}=349$ corresponding to ~17 K). According to Eq. \ref{eq:conversion_b1}, this results in a $C_{\rm{P}}$ variation of $\sim12\%$, which is remarkable considering an effective temperature tuning range of $\sim$13 K.

\begin{figure*}[htb]
	\centering
	\includegraphics{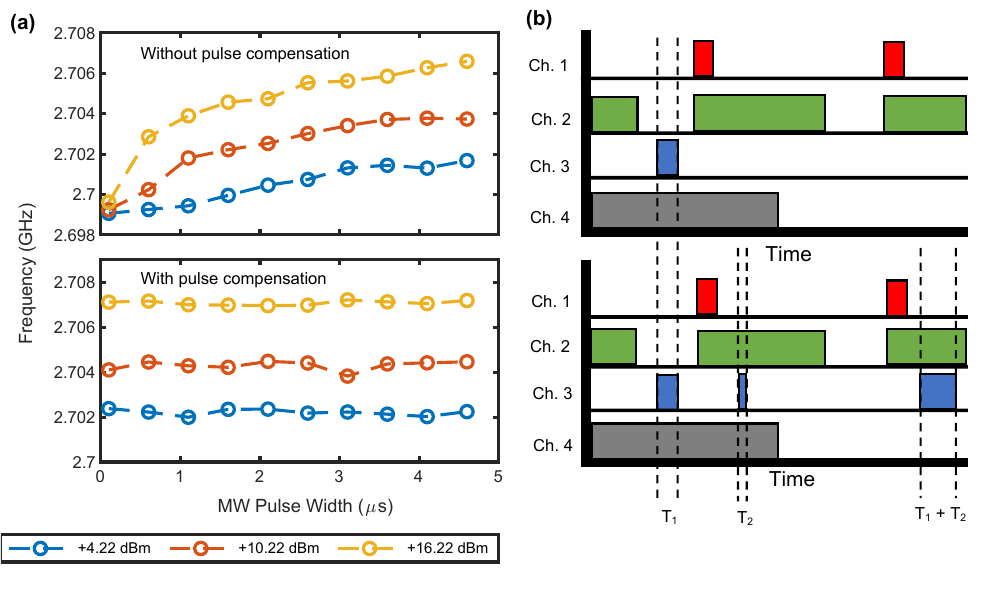}
	\caption{Dielectric resonator frequency stabilization.
	\textbf{(a)} The resonator frequency drift is likely due to heating caused by longer microwave pulses and higher microwave powers when no pulse compensation is employed. All lasers are off.
	\textbf{(b)} The resonator frequency is stabilised by keeping the total microwave duty cycle (T\textsubscript{1}+T\textsubscript{2}) constant for various microwave pulse widths (T\textsubscript{1}). In the first half cycle a measurable ODMR signal is created, while the second half cycle acts as reference for the lockin amplifier. Channel 1: Detector gating, Channel 2: 520~nm laser excitation, Channel 3: Microwave pulse, Channel 4: Lockin reference.}
	\label{fig:mwpulse_control}
\end{figure*}

The resonator is tuned continuously in the background during all experiments since the resonator otherwise exhibits frequency drifts over a short time period, which is attributed to thermal fluctuations caused by the pulsing nature of laser and microwave radiation. Additionally, we also implement a MW pulse sequence wherein we minimize the resonator frequency drift by ensuring a constant duty cycle (i.e., total pulse time) within each pulse sequence repetition (refer to Fig.~\ref{fig:mwpulse_control}). We estimate the dielectric resonator temperature to rise by 1.5 K from the base temperature of 4 K with an application of 10 dBm of pulsed microwave power. Although, it may be noted that such a large MW power is not required for measurements involving single defects due the lack of ensemble broadening effects, reducing the thermal load on the cryostat.

The optical setup and measurement principles have been adapted from Ref.~\onlinecite{Sewani2020}, with an addition of a 785~nm laser and respective optics for resonator tuning. An aspheric lens with a long working distance of 12.43~mm is used to focus the lasers beams and collect the sample emission through the topmost dielectric resonator. The use of an aspheric lens with a long working distance results in disadvantages such as chromatic aberration and low NA, which requires higher laser powers to maintain sufficient photoluminescence signal. Continuous wave excitation using the 520~nm laser corresponds to 11.5~mW of optical power. The resulting photoluminescence is measured as roughly 12~Mcps, with a 0.11\% signal fluctuation. The average optical power delivered to the sample is much lower for pulsed measurements such as ODMR, Rabi oscillations and Hahn echos, resulting in significantly lower average photoluminescence count rates.

Both the \Bzero magnetic field from the superconducting solenoid and the \Bone mode from the dielectric resonator are oriented along the z-axis, corresponding to the $[100]$ crystal direction. This leads to quenching of the NV\textsuperscript{-} centres' photoluminescence at relatively low \Bzero intensities, as the quantization axes shift away from the $\langle111\rangle$ \nv axes \cite{Tetienne2012}. A related issue is that the dielectric resonator frequency must be tuned to the ODMR frequency of the NV\textsuperscript{-} centres for meaningful measurements. A high 520 nm laser power could heat the dielectric resonator to such an extent that the \Bzero field must be increased accordingly to shift the ODMR frequency closer to the resonator frequency. A trade-off between laser power and magnetic field quenching must be made so as to optimise the signal to noise ratio for spin signal measurements. Therefore, a magnetic field of roughly 5~mT is used for the spin measurements which allows for sufficient spin signal detection without significant quenching.

A summarised list of instruments used and their applications are listed in Table \ref{tab:instruments}.

\begin{table}[htb]
    \centering
    \begin{tabular}{ll}
     \hline
     \hline
     \textbf{Instrument} & \textbf{Application} \\ 
      \hline
     \shortstack{ Bluefors LD \\ Dilution Refrigerator} & \shortstack{\\Cryostat (4 K using only\\ pulse tube cooling)} \\
     \hline
     Keysight N5183B MXG & Microwave source\\  
     \hline
     Agilent 8753ES  &  Vector network analyzer\\ 
     \hline
     CentricRF CF2040  & Microwave circulator\\
     \hline
     Spincore PulseBlasterESR-PRO  & Pulse generator \\
    \hline
     \shortstack{\\Minicircuits \\ ZACS622-100WSX+}    & Microwave power combiner\\ 
    \hline
     Minicircuits ZHL-5W-63-S+ & Microwave power amplifier\\ 
    \hline
     SRS SR830 & Lock-in amplifier\\ 
     \hline
     \shortstack{\\Attocube ANSxy100lr \\ + ANSz100std} & Piezoelectric positioner\\
    \hline
     Excelitas SPCM-ARQH-10-FC & Single photon detector\\
    \hline
     Thorlabs LP520-SF15 & 520~nm laser diode\\
    \hline
    \shortstack{\\Thorlabs FPL785P-200 \\ + CLD1015} & \shortstack{785 nm laser diode \\ + laser driver}\\
    \hline
    \hline

    \end{tabular}
    \caption{List of all relevant instruments/apparatus used.}
    \label{tab:instruments}
\end{table}

\subsection{CST magnetic field simulation}\label{cst_sim}
\begin{figure*}[hbt]
	\centering
	\includegraphics{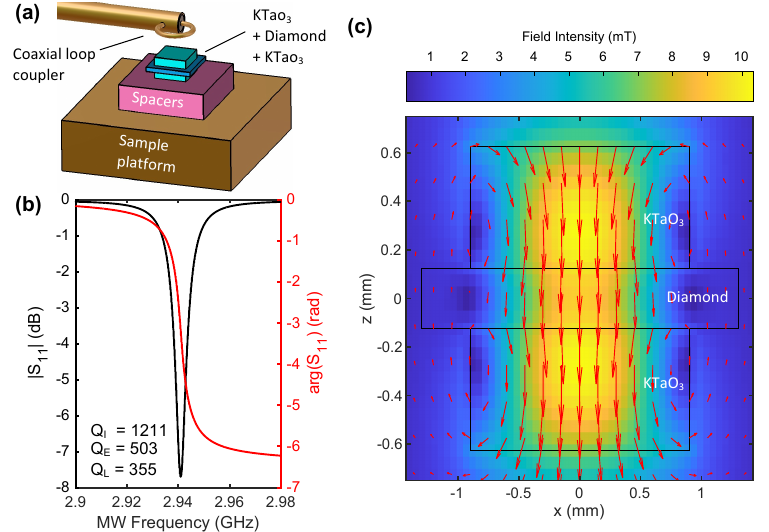}
	\caption{Magnetic field simulation. \textbf{(a)} Simulation model showing the loop coupler and relevant samples stacked on a conductive platform. \textbf{(b)} Simulated S\textsubscript{11} magnitude showing fitted quality factors similar to measured values in Fig. \ref{fig:sample_setup_diagram}(c). 
	\textbf{(c)} The simulated magnetic field distribution within the relevant sample volume with 1 W of applied microwave power at the loop coupler.}
	\label{fig:all_hfield}
\end{figure*}

The coupling between the loop coupler and the sample stack is simulated in CST Microwave Studio. The simulation results are plotted in Fig. \ref{fig:all_hfield}. The relevant simulation parameters are given in Table \ref{tab:cst_sim_params}. The simulation parameters are adjusted to match the experimental setup conditions and measured coupling.

The mode volume of the dielectric resonator is calculated from the simulated H-field (A/m) of the structure. The relevant expression is given as $V_{m} =\int_{V}\frac{|H^{2}(r)|}{|H^{2}_{\rm max}(r)|} dV$ \cite{Breeze2018}. The integral is evaluated over the H-field generated by the structure and $H_{\rm max}(r)$ represents the maximum H-field intensity found within the structure.

\begin{table}[h]
    \centering
    \begin{tabular}{lcl}
    \hline
    \hline
     \textbf{Parameter} & \textbf{Value} & \textbf{Description} \\ 
     \hline
     d1 & 1 mm & Coax inner conductor diameter \\
     d2 & 1.25 mm & Separation between top \\
     & & \ktao resonator and loop coupler \\
    %  \hline
     $\epsilon_{rk}$ & 3300 & \ktao relative permitivitty\\
     $\epsilon_{rd}$ & 5.68 & Diamond relative permitivitty\\
     $\epsilon_{rs}$ & 9.4 & Sapphire relative permitivitty\\
    %  \hline
    Tan $\delta_{k}$ & $2.5\cdot10^{-5}$  &  \ktao dielectric loss tangent \\
    %  \hline
    Tan $\delta_{d1}$ & 1.25  &  \nv Diamond dielectric \\
    & & loss tangent\\
    Tan $\delta_{d2}$ & $1\cdot10^{-4}$  & Electronic grade Diamond  \\
    & & dielectric loss tangent\\
    Tan $\delta_{s}$ & $4\cdot10^{-4}$  & Sapphire dielectric loss tangent \\
    %  \hline
    $\rho_{c}$ & $2\cdot10^{8}$ & platform conductivity\\
     \hline
     \hline

    \end{tabular}
    \caption{Summary of relevant parameters used in the CST simulation model.}
    \label{tab:cst_sim_params}
\end{table}

\subsection{Quality factor comparison}
\label{q_factor_comparison}
We observed that the quality factor of the coupled dielectric resonator was generally lower than expected. Upon further investigation we determined that it was in fact the \nv diamond sample that was responsible for the low quality factors. The results of reference measurements without diamond samples and with electronic grade diamond are summarised in Table~\ref{tab:diamond_comparison}. The \nv diamond sample we used in the measurements contains boron and nitrogen impurities, which may be the cause for the increased loss tangents and the low loaded quality factor. Using a CST simulation model (see Section \ref{cst_sim}) we have estimated the loss tangents of the electronic grade diamond to be $\approx 1\cdot10^{-4}$, which is comparable to literature values for the relevant temperature and frequency range. \cite{parshin2005dielectric,floch2011}. Similarly the loss tangent of the of the \nv diamond sample is found to be 1.25, which is four orders of magnitude higher.

\begin{table}[h]
    \centering
    \begin{tabular}{|c|c|}
        \hline
        & \textbf{Single \ktao}\\
        \hline
        $Q_{\rm I}$  & $61,240 \pm 330$\\
        $Q_{\rm L}$  & $31,220 \pm 110$\\
        $Q_{\rm E}$  & $63,690 \pm 150$\\
        
        \hline
        & \textbf{Single \ktao + \nv CVD diamond}\\
        \hline
        $Q_{\rm I}$  & $3,360 \pm 80$\\
        $Q_{\rm L}$  & $2,990 \pm 70$\\ 
        $Q_{\rm E}$  & $27,840 \pm 460$\\
        
        \hline
        & \textbf{\ktao sandwich + \nv CVD diamond}\\
        \hline
        $Q_{\rm I}$  & $2,340 \pm 50$\\
        $Q_{\rm L}$  & $2,060 \pm 40$\\
        $Q_{\rm E}$  & $17,050 \pm 250$\\
        
        \hline
        & \textbf{Single \ktao + electronic grade}\\
        & \textbf{CVD diamond}\\
        \hline
        $Q_{\rm I}$  & $\hspace{9pt}57,400 \pm 1,000$\\
        $Q_{\rm L}$  & $46,400 \pm 800$\\
        $Q_{\rm E}$  & $\hspace{4pt}243,000 \pm 3,000$\\
        \hline
    \end{tabular}
    \caption{Comparison of Q factors of various samples and arrangements. $Q_{\rm I}$,$Q_{\rm L}$, $Q_{\rm E}$ refer to internal, loaded and external coupling respectively.}
    \label{tab:diamond_comparison}
\end{table}
%%%%%%%%%%%%%%%%%%%%%%%%%%%%%%%%%%
% \clearpage
\subsection{\Bone conversion factor}
\label{Supp_B1Conv}
To calculate a fair estimate for the microwave power to \Bone conversion factor $C_{\rm P}$, we have taken into account the four possible orientations of the NV\textsuperscript{-} axes, the microwave line loss, insertion loss of the various microwave components used, as well the power reflected from the coaxial loop coupler. Our microwave losses, including insertion losses, reflection as well as the amplifier gain are quantified in Table \ref{tab:mw_loss}.

There is some microwave power that is reflected from the loop coupler. We can compensate for this reflected power from the loop coupler by measuring the S\textsubscript{11} (dB) parameter using the VNA. The appropriate `gain' $G_{\rm c}$ in dB is given by
\begin{equation}
G_{\rm c} = 10\cdot\log_{10}{(1-(10^{\frac{S\textsubscript{11}}{20}})^{2}),}
\label{eq:compensation}
\end{equation}
and represents the transmission loss. This value is simply added to the microwave line gains and losses shown in Table~\ref{tab:mw_loss} (excluding the reflection loss). This value can also be used to compensate the measured Rabi frequencies, due to changes in reflection loss as a result of changed coupling positions of the dielectric resonator to the loop coupler.

\begin{table}[htbp]
    \centering
    \begin{tabular}{lcl}
    \hline
    \hline
     \textbf{Component} & \textbf{Gain/Loss (dB)} & \textbf{Description} \\ 
     \hline
    Room temperature & -33.06  & Attenuators \\
    cables \& attenuators & & \& cable loss\\
    %  \hline
     MW power amplifier & +45.41  &  Amplifier gain\\
    %  \hline
     MW power combiner & -3.36  &  Insertion loss\\
    %  \hline
     Cryostat MW line & -3.34  &  2.967 GHz, 4 K \\
    %  \hline
     MW circulator & -0.60  & Insertion loss.\\
    %  \hline
     Reflection loss & -0.83  &  S\textsubscript{11}$=$-7.60 dB \\
     & & using Eq. (\ref{eq:compensation})\\
     
     \hline
     Total & +4.22 & \\
     \hline
     \hline

    \end{tabular}
    \caption{Table of all relevant microwave losses and gains in the microwave line.}
    \label{tab:mw_loss}
\end{table}

The \Bone field is known to be strongly oriented in the $[100]$ direction. An approximation can be made that all four \nv orientations thus experience the \Bone field at the same angle, corresponding to roughly $\alpha = 54.7^{\circ}$. The measured Rabi frequency is given by the relationship $\Omega_{R} = \frac{1}{\sqrt{2}}\gamma_{e}B_{1\perp}$, where $B_{1\perp}$ is the perpendicular (to the NV\textsuperscript{-} axes) component of the total $B_1$ field driving the spins. The total \Bone magnitude is thus given as $B_1 = \frac{B_{1\perp}}{\sin{\alpha}}$\cite{Eisenach2018}.

\subsection{Rabi coherence simulation}\label{Supp_Rabi}

Simulation of the Rabi oscillation decay times requires the following parameters:

\begin{itemize}
    \itemsep0em 
    \item Static magnetic field vector $B_0$.
    \item Simulated AC magnetic field ($B_1$) vectors and frequency.
    \item Gaussian ensemble broadening.
    \item Hyperfine coupling (MHz).
    \item Laser intensity within the sample volume.
\end{itemize}

The \Bzero field vector and the ensemble broadening parameters allow us to calculate the Rabi oscillation signal contributions arising from detuned spin transitions. The OMDR transition frequencies themselves can be calculated from solving the relevant eigenvalues of the \nv Hamiltonian similar to Ref.~\onlinecite{Sewani2020}. Though Rabi oscillations can be calculated directly from the Hamiltonian as well, it is preferable to use the analytical Rabi formula in the rotating frame given in Eq.~\ref{eq:rabi} to reduce the computation time.

\begin{equation}
    P_{N\ket{\uparrow}} = \frac{\omega^{2}_{1,N}}{\Delta\omega^{2}+\omega^{2}_{1,N}}\sin^{2}{\left(\frac{\sqrt{\Delta\omega^{2}+\omega^{2}_{1,N}}}{2} t \right)} 
\label{eq:rabi}
\end{equation}

\begin{figure*}[hbt]
	\centering
	\includegraphics{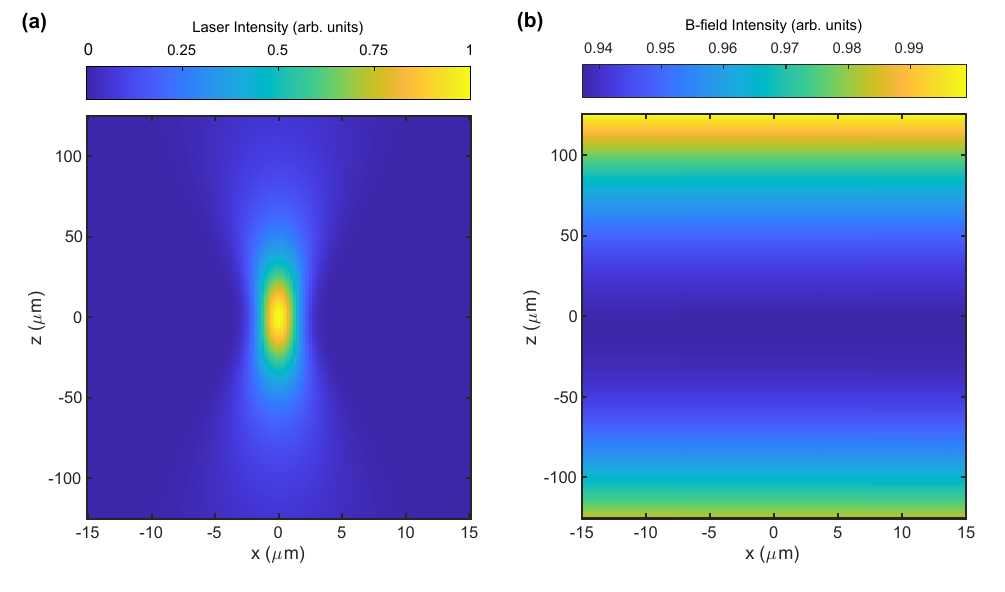}
	\caption{Laser excitation and magnetic field simulations.
	\textbf{(a)} Cross section plot of the 520~nm laser intensity model within the signal collection volume showing a minimum beam waist of $4.6$~$\mu$m at the focus. 
	\textbf{(b)} Plot of the simulated B-field magnitude within the signal collection volume. The B-field intensity has been re-scaled.}
	\label{fig:laser_bfield}
\end{figure*}

The \Bone distribution within the signal collection volume is represented as a function $\omega$\textsubscript{1,\textit{N}}(x,y,z) which can be obtained from an EM simulation of the structure, and is plotted in Fig.~\ref{fig:laser_bfield}(b). This $\omega$\textsubscript{1,\textit{N}}(x,y,z) is calculated for each NV\textsuperscript{-} orientation labelled by the subscript $N$. The \Bone data is scaled according to the Rabi frequency measured for each MW power applied. Further, as we know that the component of the \Bone field perpendicular to the NV\textsuperscript{-} axis is responsible for the oscillations, we must scale the \Bone field by a factor of $\frac{\sin{(\theta)}}{\sin{(54.7^{\circ})}} \Omega_{\rm R}$ (approximation). Here $\theta$ represents the angle (in degrees) between the NV\textsuperscript{-} axes and the magnetic field vector $B_1$. 

The signal collection volume is effectively given by a simple gaussian laser intensity function within the diamond sample volume: 
\begin{equation}
f(x,y,z) = \left(\frac{W_{0}}{w_{z}}\right)^{2}  e^{-2\left(\frac{x-\mu_{x}}{w_z}\right)^{2}}  e^{-2\left(\frac{y-\mu_{y}}{w_z}\right)^{2}}
\label{laser_intensity}
\end{equation}
\begin{equation}
W_0 = \frac{4M^2\lambda{}f_l}{2\pi{}D},
\end{equation}
\begin{equation}
w_z = W_0\sqrt{1+\left({\frac{z\lambda}{\pi{}W_{0}^2}}\right)^{2}}.
\end{equation}

The parameters of this function are summarized in Table~\ref{tab:laser_params}. 

\begin{table}[h]
    \centering
    \begin{tabular}{lcl}
    \hline
    \hline
     \textbf{Parameter} & \textbf{Value} & \textbf{Description} \\ 
     \hline
     D & 2.2 mm & Collimated beam diameter \\
    %  \hline
     $\lambda$ & 520 nm  & Laser wavelength \\
    %  \hline
      $f_{l}$ & 15.29 mm  &  Lens focal length\\
    %  \hline
     M & 1  &  Beam ellipticity\\
    %  \hline
     2$W_0$ & 4.6 $\mu$m & Focused spot diameter\\
    %  \hline
     $\mu{}_{x,y}$ & 0 mm &  Beam displacement\\
     \hline
     \hline

    \end{tabular}
    \caption{Description of parameters used in the laser intensity function.}
    \label{tab:laser_params}
\end{table}

The excitation laser spot size is calculated to be 4.6~$\mu$m, given the lens focal length of 15.29~mm and a collimated beam diameter of 2.2~mm. Then, using Eqs.~\ref{eq:rabi} and \ref{laser_intensity} we derive the total Rabi signal as a function of time and summed over the $N = 4$ possible \nv  orientations as given in Eq.~\ref{total_signal}:
\begin{equation}
% \small
\Omega_{R}(t) = \sum_{N = 1}^{4}\int_{V}^{}f(x,y,z)P_{N\ket{\uparrow}}(\omega_{1}(x,y,z),t)\,dV
\label{total_signal}
\end{equation}

The limits of the integration represent the bounds of the sample volume of interest as illuminated by the laser. It is the decay time constant of this signal that corresponds to the T\textsubscript{2}\textsuperscript{Rabi} coherence time.

The non-linear data fitting is accomplished using Matlab's global optimization toolbox. Rough fitting is done with the `PatternSearch' algorithm, and finer fits are accomplished by the `nlinfit' function which can also provide fit parameter confidence intervals. The list of fit parameters and their values are given in Table~\ref{tab:opt_sim_params}.

\begin{table}[h]
    \centering
    \begin{tabular}{ll}
    \hline
     \hline
     \textbf{Parameter} & \textbf{Value} \\ 
     \hline
      Ensemble broadening & $1.868\pm0.054$ MHz \\
    %  \hline
      $B_{1}$ angle & $3.723^{\circ}\pm0.012^{\circ}$ \\
    %  \hline
      $B_{0}$ angle & $0.000 ^{\circ}$ \\
    %  \hline
     $B_{0}$ magnitude & $5.031\pm0.023$ mT \\
    %   \hline
     $D_{z}$ (zero field splitting) & 2.878 GHz \\
    %  \hline
     Hyperfine & 2.15 MHz \\
     \hline
    \hline

    \end{tabular}
    \caption{Optimised simulation parameters for simulation shown in Fig.~\ref{fig:rabi_power_decay_simulation}(a) in the main text. All angles are given w.r.t the $[100]$ crystallographic direction.}
    \label{tab:opt_sim_params}
\end{table}

In our simulations, we find that we require approximately $3.7^{\circ}$ of tilt in the \Bone vector for the simulation results to match the measurement data. We partly attribute this tilt to imprecise sample mounting and a crystallographic miscut of the diamond sample. The manufacturer's specification of the crystallographic orientation miscut tolerance is given as $\pm 3^{\circ}$\cite{E6}.

\begin{figure*}[htb]
\centering
    \includegraphics{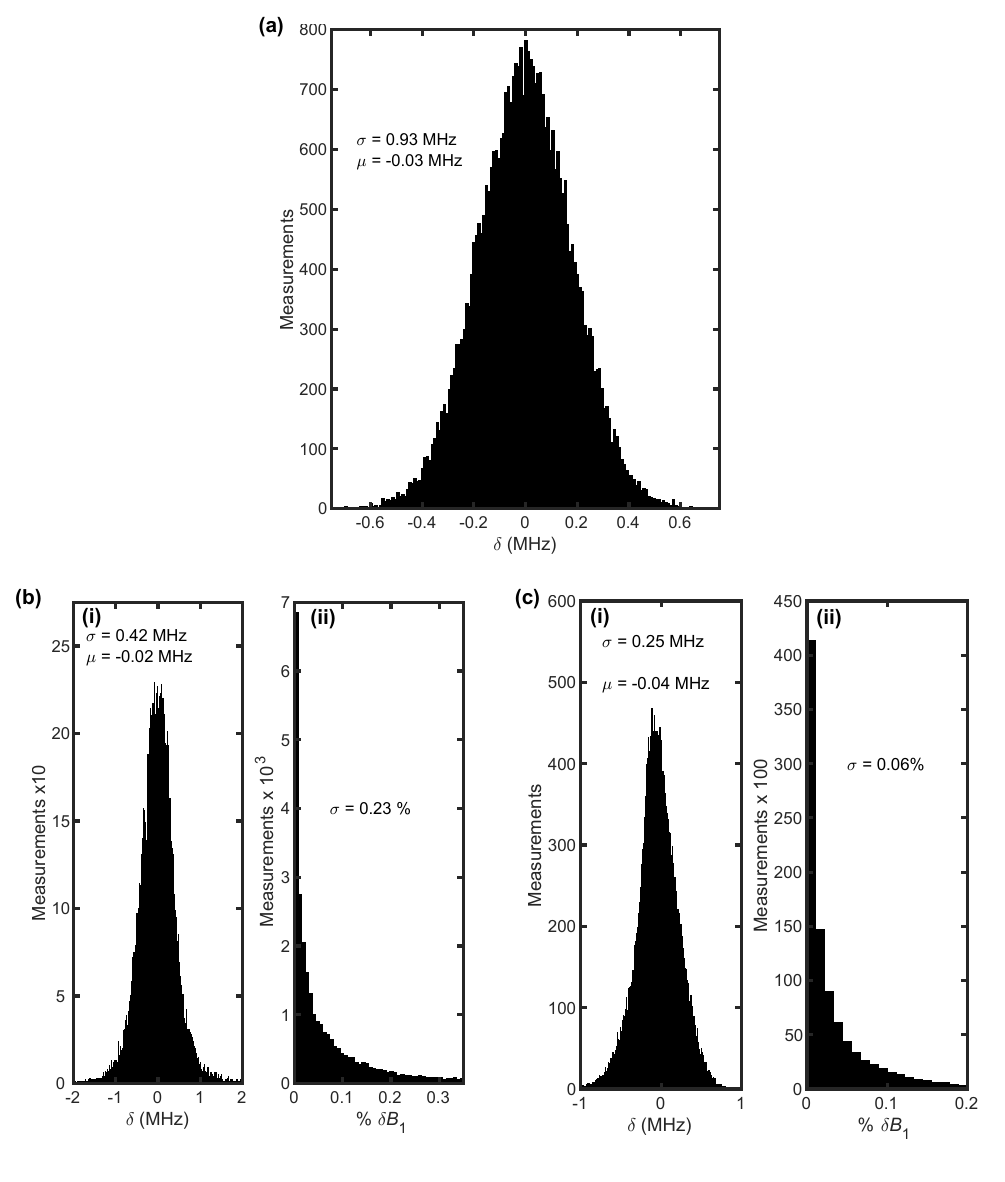}
    \caption{Dielectric resonator frequency stability.
    \textbf{(a)} Resonator deviation from 2.967 GHz as recorded while implementing continuous resonator tuning during the ODMR measurement shown in Fig.~\ref{fig:sample_setup_diagram}(d) in the main text. 
    \textbf{(b)} Histogram of resonator frequency deviation from the set point, accumulated throughout the measurement corresponding to Fig.~\ref{fig:rabi_power_decay_simulation} in the main text. A Corresponding histogram data shows the distribution in terms of percentage change of the Rabi frequencies ($\delta B_1$) as a result of resonator deviation. The calculation of $\delta$\Bone uses measured data shown in Fig.~\ref{fig:detuning_chevron} in the main text.
    \textbf{(c)} Combined histogram of all measured resonator deviation values from the detuning setpoints used for the measurement shown in Fig.~\ref{fig:detuning_chevron} in the main text.
    }
    \label{fig:histograms}
\end{figure*}

While there is some instability in the resonator frequency, despite our control/tuning mechanism (see Fig.~\ref{fig:histograms}), we find that such instability does not significantly impact the coherence times of our Rabi oscillations.

\begin{figure*}[htb]
	\centering
	\includegraphics{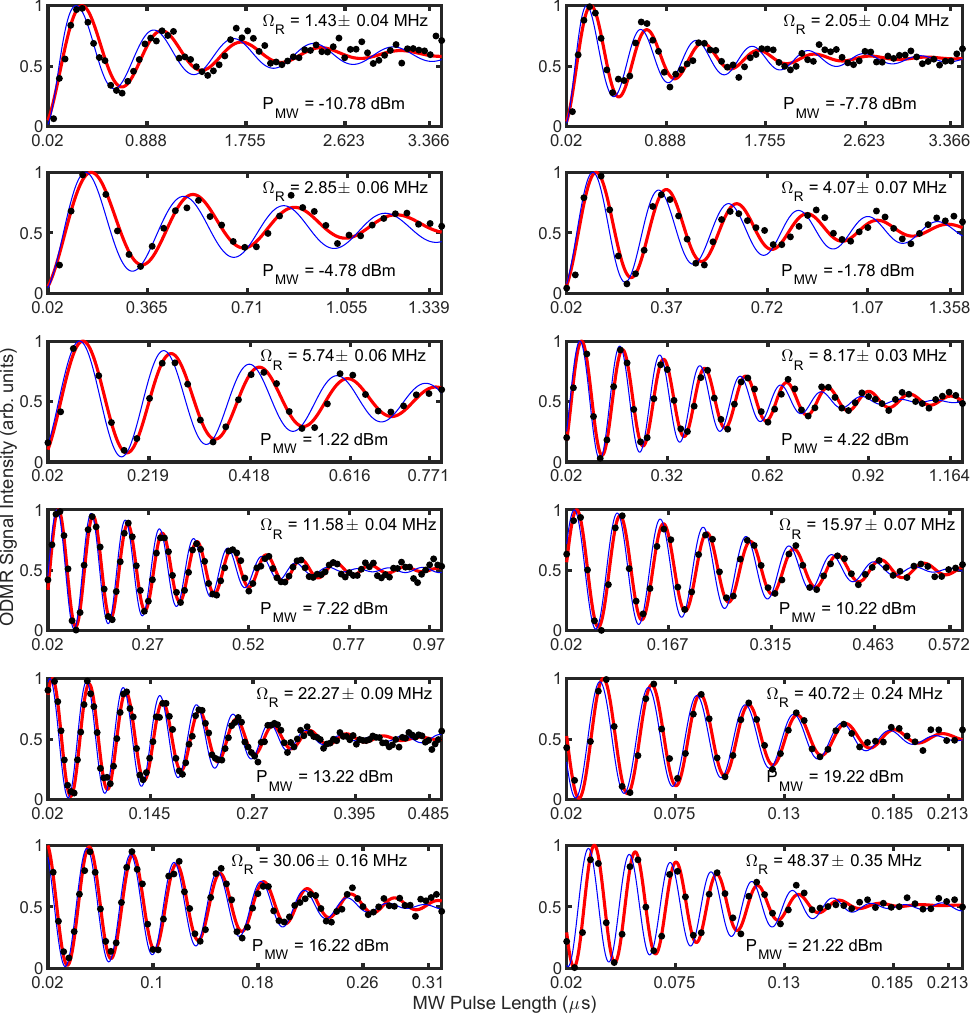}
	\caption{Collection of measured (red) and simulated (blue) Rabi signals for assorted applied microwave powers. The measured data are fit to the function $f(t) = e^{-(t/\tau)^{n}}\sin(2\pi\Omega_{\rm R}t+p)+c$.}
	\label{fig:rabi_collection}
\end{figure*}

\begin{figure*}[htb]
	\centering
	\includegraphics{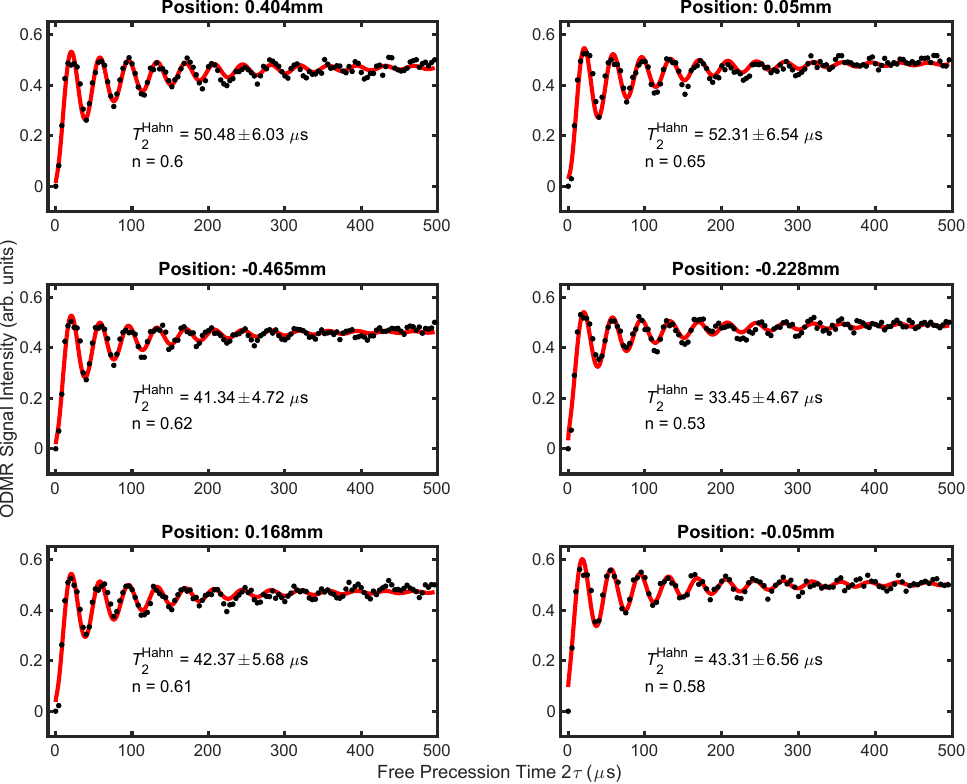}
	\caption{
	Collection of measured and fitted $\tau^{2}$ Hahn echo signals for various positions along the x-axis of the resonator. The data points at the end of the measured signals have been scaled to 0.5 to represent the spins entering into a mixed state. The fit function used is $f(t) = e^{-(t/\tau)^{n}}(a_{1}+a_{2}(\sin^{2}(0.5\omega t+p)))+c$.}
	\label{fig:hahn_analysis_plot_all}
\end{figure*}

\clearpage
\bibliography{library}
\end{document}